# Opening discussion on banking sector risk exposures and vulnerabilities from virtual currencies:
*An operational risk perspective.*

Gareth W. Peters[1,2,3], Ariane Chapelle[4,5] and Efstathios Panayi[6]

1. Department of Statistical Science, University College London, London, UK
2. Associate Member, Oxford Mann Institute (OMI), Oxford University, Oxford, UK
3. Associate Member, Systemic Risk Centre, London School of Economics, London, UK
4. Honorary Reader, University College London, London, UK
5. Chappelle Consulting (chappelleconsulting.com)
6. Financial Computing Centre, Department of Computer Science, University College London, London, UK

## Abstract

We develop the first basic Operational Risk perspective on key risk management issues associated with the development of new forms of electronic currency in the real economy. In particular, we focus on understanding the development of new risks types and the evolution of current risk types as new components of financial institutions arise to cater for an increasing demand for electronic money, micro-payment systems, Virtual money and cryptographic (Crypto) currencies.

In particular, this paper proposes a framework of risk identification and assessment applied to Virtual and Crypto currencies from a banking regulation perspective. In doing so, it addresses the topical issues of understanding important key Operational Risk vulnerabilities and exposure risk drivers under the framework of the Basel II/III banking regulation, specifically associated with Virtual and Crypto currencies. This is critical to consider should such alternative currencies continue to grow in utilisation to the point that they enter into the banking sector, through commercial banks and financial institutions who are beginning to contemplate their recognition in terms of deposits, transactions and exchangeability for fiat currencies.

We highlight how some of the features of Virtual and Crypto currencies are important drivers of Operational Risk, posing both management and regulatory challenges that must start to be considered and addressed both by regulators, central banks and security exchanges. In this paper we focus purely on the Operational Risk perspective of banks operating in an environment where such 'electronic' Virtual currencies are available. Some aspects of this discussion are directly relevant now, whilst others can be understood as discussions to raise awareness of issues in Operational Risk that will arise as Virtual currency start to interact more widely in the real economy.

We propose a structure of risk analysis starting with the exposures and vulnerabilities of virtual and crypto currencies as the drivers of operational risk for these new means of exchange. Then by using risk drivers, our approach allows us to highlight the sources of possible adverse consequences, when using or generating virtual and crypto currencies. These are then mapped into the risks associated to the Basel categories, providing an easier view of regulatory response, and better mitigation techniques. In addition, this will help identify and address the root causes of the Operational Risks associated with virtual and crypto currencies, rather than just presenting their symptoms.

# 1. Introduction

The emergence of electronic money in some form or another, such as for instance micro payment systems, is not a new concept for banking regulation. Indeed, it has been discussed for several decades in a generic way as noted by [Drehmann et al, 2002] where they succinctly provided a brief description of some key reasons why exactly electronic money is becoming the norm through a quote from James Gleick that we repeat due to its ability to crudely but succinctly capture the fundamental drivers for 'modernisation' of money:

*"Cash is dirty ... Cash is heavy ... Cash is inequitable ... Cash is quaint, technologically speaking .. Cash is expensive ... Cash is obsolete"* (Gleick, 1996)

If one adopts the stance reflected in this quote and by [Drehmann et al, 2002], also evidenced by the growing interest in virtual, crypto and micro-payment systems, as reflected by current market caps for the leading virtual currencies exceeding 10mil USD on 1/9/2014[1]: Bitcoin (market cap $ 6,250,485,561); Litecoin (market cap $ 152,083,510) ); Ripple (market cap $ 143,742,338); BitSharesX (market cap $ 54,416,797); Nxt (market cap $ 26,478,823); Peercoin (market cap $ 17,045,108); Dogecoin (market cap $ 11,787,931); Darkcoin (market cap $ 10,733,081); and Namecoin (market cap $ 10,296,430). Then, given the increasing volumes, exchange trading activity and growth in these virtual market places over the last few years, it is not inconceivable that one would expect that some form of virtual currencies will eventually enter into the banking sectors and the real economy. Indeed, some would argue this has already occurred, at least in some jurisdictions. Following up on this point, we also note that the world now has a larger number of virtual currencies than the total of the fiat recognised currencies in different parts of the globe, of which there are 467 which are exchange traded in USD as of August, 2014[2]. In addition, there are a wide array of exchanges and market places arising primarily to facilitate the increasing activity in trading such Virtual currency and Crypto currencies, such as: Coinbase.com; Cryptsy.com; Bter.com; and Coins-e.com[3]. From this vast array of virtual currencies in circulation and being traded on bourses (exchanges) and markets across the internet, none of them carry an official stamp from the government or banking regulator from any of the countries that they are being bought, sold, exchanged or bartered within the real economy.

Whilst there is a history of studying the role and impact of electronic money in banking networks from many perspectives including regulation and risk, we believe there is an interesting new dimension to be added to this literature. This involves the discussion on non-fiat denominated electronic Virtual and Crypto currencies which are currently operating beyond the realm of central banks and regulatory authorities. We advocate in this paper that a more concerted effort be made to clearly study and develop policy relating to the utilisation of such non-fiat Virtual and Crypto currencies in order for banking regulators and financial sectors to be prepared for the financial risk management features that will arise specifically in the context of banking networks admitting transactions, deposits and exchange in these new non-standard currencies. In addition, we take the stance, backed up with increasing empirical evidence and market capitalizations by certain Virtual and Crypto currencies, that particular forms of Crypto currency will eventually become part of the mainstream financial banking networks, in the not too distant future. We therefore advocate a proactive response, developing in this paper an initial dialogue and carefully considered decomposition of the fundamental features such Virtual and Crypto currencies present as vulnerabilities and risk exposures which may act as drivers for certain types of financial risks and subsequent losses.

---

[1] http://coinmarketcap.com/
[2] Virtual currency exchange: http://coinmarketcap.com/currencies/views/all/
3 A comprehensive list of such currency exchanges currently available: http://planetbtc.com/complete-list-of-Bitcoin-exchanges/

The particular focus we present in this paper is to raise awareness for regulators, banks and financial institutions, who may be considering the possibility of admitting Virtual and Crypto currencies into banking networks, of the inherent financial risks that are specific to Virtual and Crypto currencies. Primarily our focus in this paper is on the sources of Operational Risk, as described by the Basel II/III banking regulations, which would arise, should Virtual or Crypto currencies be allowed to be accepted as transactions, deposits or exchanges within standard banking environments and networks. The emphasis on Operational Risk is not incidental, it is selected as we believe it will present as one of the fundamental sources of financial risk that will specifically be of concern for Virtual and Crypto currencies in banking environments. The remainder of the paper will clearly outline why we hold this belief.

In this paper, we attempt to begin to isolate the main drivers of Operational Risk's embedded in virtual and crypto currencies, to highlight some of their consequences in terms of operational events and losses for stakeholders of virtual currencies. In particular we will focus attention in this paper on virtual and cryptographic currency, for discussions on other related aspects such as micro-payment systems see [Párhonyi et al., 2005] and the references therein. We consider the impact on key aspects of Operational Risk that will need to be adjusted and developed to tackle the different nature of risk profiles such electronic currencies will present compared to current best practice in Operational Risk management for fiat money. This is of relevance, as currently to date the papers that begin to appear on Virtual currencies and Crypto currencies regulation have not considered in particular the impact on banking regulation such as Basel II/III.

Whilst the developments of such currencies is rapidly progressing from the computer sciences, software engineering and mathematical modelling perspective, there is still a dearth of knowledge relating to the risk management and regulation components of Virtual and Crypto currencies. This is however starting to be addressed, initially by regulators and central banks at least releasing statements in an initial response or attempt to classify Virtual and Crypto currencies within the real economy. Therefore, we also believe our paper will be of benefit to regulators and the banking sector, to help further understand important aspects of Virtual and Crypto currencies, not well understood by the greater population as they begin to explore the possibilities offered by such non-standard currencies. We believe that currently the lack of understanding of the core features of such Virtual and Crypto currencies has made regulatory response difficult to develop. This, when combined with price instability displayed in early exchange rates for Virtual and Crypto currencies affects the consumers' confidence in these new currencies which can have significant implications on their value, destabilising the micro-economies arising around them. We believe a careful consideration of the features and risk profiles of such Virtual and Crypto currency networks will help to resolve some aspects of these uncertainties and provide a sound basis for informed decision making by regulators and consumers alike.

Therefore, taking the stance that some form of virtual currencies will eventually enter into the banking sectors, we aim to bring a greater awareness to particular financial risk analysis that ought to be considered for Virtual and Crypto currencies. This awareness will begin to arise as increasingly more Operational Risk loss events arise, specifically related directly or indirectly to Virtual and Crypto currencies. It is useful at this stage to recall the basic notion of Operational Risk according to Basel II/III where it is defined at a high level as follows:

*Operational risk is the risk of loss resulting from inadequate or failed internal processes, people, and systems or from external events. (Basel II, Solvency II)*

It is clear just from this very basic definition of Operational Risk that as Virtual and Crypto currencies become more active in the real economy and banking systems, they will generate significant Operational Risk challenges. Still in infancy, Virtual currencies and their subset, the Crypto currencies, face currently the double jeopardy of generating significant operational risks whilst suffering from immature operational risk management framework and practices. Massive losses arising from such operational risk exposures are already starting to materialise in a very public manner such as in the very recent case of Mt Gox's losses of around 750,000 Bitcoins to hackers, a net worth of more than $470 million USD at the time[4]. For a detailed account of all modelling aspects of Operational Risk we refer the interested reader to recent book length discussions in [Cruz, Peters and Shevchenko, 2014] and [Peters and Shevchenko, 2014b]. We do not enter into the specific risk models in this paper, instead focusing on discussing core features of the Operational Risk vulnerabilities and exposures that will eventually need to be modelled.

This paper is organised as follows. Section 2 introduces a background on virtual and crypto currencies including a brief description of the different forms: electronic money, Virtual currency and Crypto currency. Section 3 provides an overview of regulatory responses to Virtual and Crypto currencies. Section 4 develops a taxonomy of these currencies and Section 5 develops an analysis of the Operational Risk vulnerabilities and exposures arising from Crypto currencies should they enter into financial or banking sectors. Section 6 concludes the analysis.

## 2. Background on Virtual and Crypto Currencies

### 2.1 Emergence of Virtual and Crypto Currencies

Whilst it is clearly beyond the scope of this paper to provide a detailed and comprehensive coverage of the history and evolution of different monetary systems, it is interesting to make a brief mention of monetary forms over time. A detailed account of monetary evolution can be found in the reviews of [Grierson, 1977] and [Davies, 2005].

Paper money akin to modern banknotes emerged in the 9th century AC in China as a response to a shortage in raw materials for coinage. Then in some form or other, paper currency was used until the middle of the 15th century, when it slowly fell out of favour. In early 19th century Britain, a number of banks were able to issue banknotes, yet no effective link existed for controlling the issuance of these notes. Gold was then made the standard of value, while there was also to be some elasticity in the supply of paper money. The Bank of England was able to maintain this convertibility with very low reserves of gold until the First World War [Davies, 2005].

In the aftermath of the Second World War, the Bretton Woods agreement established the convertibility of gold into US dollars at a rate of $35 USD per ounce, and other countries fixed the exchange rate of their currencies to the US dollar. The fixed convertibility into gold remained until 1971, following which most countries started floating their currencies against the US dollar.

In addition to government backed or fixed indexation of money against the US dollar, there have also been developments in private money issuance. Besides commodity and government-backed (fiat) currencies, there have been several efforts to introduce alternative or complementary private currencies, for a number of reasons. For example, several local currencies have been introduced in the UK, with the aim of strengthening the local economy. These have largely had low volume and very localized circulation. For instance in the UK alone there is, in addition to the Great Britain Pound (GBP), the private monies in very localized circulation known as the Totnes pound[5], the Lewes pound[6], the Stroud pound[7] and the Bristol pound[8]. However, in each of these cases very limited amounts of money have been issued and none has gained widespread acceptance. In other countries such as Germany they also have some more successful versions of private money such as the Chiemgauer which started in 2003 and is named after the region in which it is locally in circulation, ie. the Prien am Chiemsee in Bavaria, Germany. This particular private money issuance has the intended goal of promoting local commerce and non-profits with its listed intended aims discussed by [Gelleri, 2009] to include: employment creation; the promotion of cultural, educational and environmental activities; the promotion of sustainability through incentives for organic food and renewable energy; the strengthening the solidarity in the region of circulation; and the stimulation of the local economy.

---

[4] http://online.wsj.com/news/articles/SB10001424052702303801304579410010379087576
[5] Totnes pound: http://www.totnespound.org/
[6] Lewes pound: http://www.thelewespound.org/
[7] Stroud pound: http://www.stroudpound.org.uk/
[8] Bristol pound: http://bristolpound.org/

In this last respect the issuance of this private money, the Chiemgauer, has been very successful to date as it continues to retain a purchasing power locally in the issuance region which is stronger than the European fiat currency the EURO. In fact there has been a turnover of over 7 million EUR in 2013[9]. For a range of discussions on the role of private money in different economies the interested reader is referred to discussions in [King, 1983] and [Dowd, 1988].

Distinct from these private monies is the emergence in the banking sector and real economy of Virtual currencies and Crypto currencies. These currencies are not localized to a particular region or country, they are not backed by any local government or private organisation and they are being circulated in the real economy on an increasingly massive scale beyond the reach of regulation, monetary policy, oversight and money supply control that has traditionally been enforced in some manner with localized private monies.

Virtual and Crypto currencies have traditionally been designed and relegated to the realms of what are often called "virtual economies" that operate within online communities such as within gaming platforms, social networking cites and online communities. Much like their fiat currency counterparts that operate in the off-line real economy, these virtual currencies can be used for the barter and purchase of virtual goods and services. Under this division between the real economy and virtual economies the Virtual currencies have been largely unregulated as it was perceived that their impact on real economies was minimal. Consequently, their understanding and regulation has been contained within guidelines or "ad-hoc" regulations of such currencies and their exchange with fiat currency, through particular gaming platform developers and a small number of private virtual currency exchanges.

However, since the establishment of Virtual and Crypto currencies purely for utilisation outside of such virtual environments, which largely occurred with the introduction of Bitcoin recently in 2008, there has been an increasing need to understand the core features and impacts such Virtual and Crypto currencies may have on the real economy, banking structures, exchange structures, regulation and oversight as well as central banking functions and monetary policy implications. Whilst the study or electronic banking networks and structures has been progressing for a few decades, see for instance early discussions on electronic banking networks in [Sifers, 1996], [Drehmann et al, 2002] and [Gup, 2003]. The focus of such works is distinct from the study and analysis being undertaken in this paper since these works focus on electronic transaction networks more aligned with micropayment systems that use traditional fiat currencies. Though related, we will be more focused on the new aspects of electronic money networks and banking structures and the risk associated with developments of non-fiat currency Operational Risk exposures in banking networks arising from Virtual and Crypto currencies in such networks.

## 2.2 Electronic money, Virtual currency and Crypto Currency

Until this point we have not made precise the distinction between electronic money, virtual currency and crypto currencies. We believe that more often than not the terms money, currency, electronic money, Crypto currency and Virtual currency are used perhaps without the careful consideration of their intended legal definitions. In this section we aim to clarify this by providing our view on the distinctions between electronic money, Virtual currency and Crypto currency, each of which we see as distinct items.

---

[9] http://www.chiemgauer.info/fileadmin/user_upload/Dateien_Verein/Chiemgauer-Statistik.pdf

The UK regulator defines electronic money as follows:

`` *Electronic money (e-money) is electronically (including magnetically) stored monetary value, represented by a claim on the issuer, which is issued on receipt of funds for the purpose of making payment transactions, and which is accepted by a person other than the electronic money issuer. Types of e-money include pre-paid cards and electronic pre-paid accounts for use online.*"

Note, in this paper we are careful not to denote Virtual currency and Crypto currency, which are digital currencies, as 'money'. The reason for this is that we will reserve the term 'money' purely for the identification of fiat based currencies which are either government or central bank backed, regulated and monitored. Most of the traditional electronic money supply is bank money held on computers. Alongside e-money, a new type of money, the virtual currency, is rapidly growing. It can be redeemed in fiat money but is not necessarily backed by such currency as a representative currency would typically be. Virtual currencies cannot be fully considered as electronic money since, although they share some attributes of electronic money, there is currently no legal founding to enforce the link between fiat physical money and Virtual currencies as there is in regulated electronic money transactions. In addition, Virtual currencies are not stored in the same unit of account as any fiat currency that would preserve their worth.

We also note that the commonly held view of the definition of money involves three key features:

1. Money should be generally accepted as a medium of exchange.;
2. Money should be a unit of account so that we can compare the costs of goods and services over time and between merchants.; and
3. Money should be a store of value that stays stable over time.

The central bank of Canada applied this criteria in a report in April 2014[10] to some Virtual and Crypto currencies and argue that such currencies fail to meet such definitions in the standard sense that fiat currencies and electronic money does, however as Virtual and Crypto currencies continue to gain acceptance and uptake in the real economy and it is indeed foreseeable that they will eventually satisfy these criteria, see further discussions on such issues in [Maurer et al, 2013] and [Zorpette, 2012].

Given these attributes, Virtual and Crypto currencies can be seen as a third form of currency type to add to the list of possible legal tender. There is no clear description in the literature yet that identifies the taxonomy of different forms of Virtual and Crypto currency currently in circulation and interacting with the real economy. In the following sections we provide a first attempt to develop such a taxonomy.

Several versions of what are now considered Virtual currencies arose out of a need to trade goods in a digital (online) environment. This could range from a game platform to other alternative virtual networks such as the ad hoc mobile networks described in [Buttyan and Hubaux, 2001] and [Zhong et al, 2003] who describe the emergence of the Nuglets, a Virtual currency used to stimulate and incentivise interaction between members of an ad hoc virtual network. However, the real evolution on a massive scale of what would be considered a class of modern successful Virtual currencies would be those issued for use in virtual economies on large online gaming platforms.

[Lehdonvirta, 2014] describes one of the earliest versions of Virtual currency, the Q coin, that successfully bridged the gap between online economy and the real economy. In 1999 Tencent Holdings launched an instant messaging service called OICQ, where each user was represented by a virtual character, and where users could spend money to customise their characters in the virtual online gaming world. Because of the lack of credit cards and payment methods to process such payments, Tencent launched a virtual currency called the Q coin, in 2002.

---

[10] http://www.bankofcanada.ca/wp-content/uploads/2014/04/Decentralize-E-Money.pdf

In the particular case of Q coins users were able to purchase new Q coin from a brick-and-mortar shop for use on the Tencent platform, so the monetary interaction between fiat currency and Q coin was initially uni-directional. However, as the success of the platform grew the virtual currency gained a wide following, and online entrepreneurs started accepting the Q coin as payment for their products or services, extending the acceptability of the Q coin to other virtual economies that also interacted with the real economy making indirectly a bi-directional interaction between fiat currency and Q coin. The wide spread uptake of this currency in some parts of China raised a number of concerns for the Chinese central bank, such as inflation considerations due to the increased money supply and the difficulty in taxing virtual currency income. They thus enacted a set of rules, which imposed limits on the issuance of such currencies and also limited their use outside the online environment in which they were originally intended for, see [Lehdonvirta, 2014] for further discussions. This is perhaps one of the earliest cases of a central bank intervention to try to regulate or control money supply of a Virtual currency.

Since this first occurrence there have been a number of Virtual currencies introduced for gaming in massively multiplayer online (MMO) or life simulation games (World of Warcraft, Diablo III, Second Life [Guo and Barnes, 2009], purchase of applications (Amazon), or games (Microsoft Xbox Live points until August 2013, Facebook credits until 2012).

Even before Virtual currencies were adopted by such online platforms, there had been a number of proposals in the cryptography literature for digital payments systems using new currencies see for instance the works of [Chaum et al, 1990], [Brands, 1995], [Frankel et al, 1998] and [Wang and Zhang, 2001]. The ideas from these works failed to achieve widespread acceptance in practical use, perhaps due to the complex security mechanisms they adopted and the inability under such approaches to easily scale to volumes required in real economies which could severely limit uptake and would limit scalability as discussed in [Simplot-Ryl et al, 2009].

In 2008 [Nakamoto, 2008] introduced both a new payments system and a new Virtual currency that was decentralised and became known as the Bitcoin. It appealed to many groups due to the nature of its creation and lack of central authority control which was unlike the vast majority of the Virtual currencies mentioned above. One of the goals of the Bitcoin protocol was to reduce online transaction costs by making financial intermediaries unnecessary, enabling direct electronic payment between individuals. In the electronic payments nomenclature, systems requiring payment verification from a broker are termed `online' payment systems, while in offline schemes (like Bitcoin) verification is not required at the point of payment. For a survey of the architectures of such schemes, see [Simplot-Ryl et al, 2009].

It is interesting to note, that [Valdes-Benavides et al, 2014] argue that for the uptake of Virtual and Crypto currencies to occur on a large scale, then three basic elements or requirements should be present if it is to be accepted in the real economy and by the public en masse: they should be low cost; they should provide reliable security; and they should offer a degree of privacy in transactions. They particularly emphasise the worth of Virtual and Crypto currencies with regard to low costs of micro-payment transactions and argue that this inherent feature will continue to drive the interest in introducing such currencies into the real economy and banking sectors.

**Crypto currency:** at this point we will begin to be more precise regarding the distinction between Virtual currencies and Crypto currencies. We will consider Crypto currencies to be a specialised sub-class of Virtual currencies with several important distinguishing features from other Virtual currencies. We will spend the remainder of the paper discussing this particular sub-class since empirical evidence supports the believe that they are the most likely candidates to be adopted in banking networks and considered as alternative legal tender compared to fiat money, in future. Therefore a clear understanding of the differences, that particularly make Crypto currencies distinct, from other forms of Virtual currency, will be important to elaborate if one is to develop an understanding of the Operational Risk inherent in incorporation of Crypto currencies into banking systems. In general we will treat Crypto currencies as a type of digital 'token' that relies on cryptography for chaining together what are known as digital signatures representing token transfers, a reliance on processing via a peer-to-peer networking system architecture and the inherent decentralization that comes from its dispersed and non-centrally governed design and protocols, see specific details in for instance [Barber et. al., 2012], [Brezo and Bringas, 2012] and [Mitsuru et.al., 2014] and the references therein.

As noted in [Mitsuru et. al., 2014] the following features define the crypto currency Bitcoin, but such features described below also have commonality with many other Crypto currencies. In particular several of the items described below are shared in some manner in the general design protocols and features of numerous Crypto currencies.

1. A central authority is not responsible for issuance or monetary policy such as money supply controls. In fact, all rules (protocols) regarding the currency design, creation and transaction are publicly available knowledge.
2. A time stamp procedure is developed to verify via the use of a peer-to-peer network whether a transaction in particular bitcoins, such as the transfer of bitcoins from one Bitcoin address to another Bitcoin address has been performed. This peer-to-peer network is not centrally controlled and anyone may join as a member. The network is also used to ensure for instance that a user has not committed the equivalent of cheque kiting with Crypto currency, i.e. double spending the currency prior to verification.
3. Crypto currencies are usually introduced into the economy gradually, and for some of the major Crypto currencies, (including Bitcoin and Litecoin), the total number of the currency units to be introduced is fixed. This is approximately 21 million coins for Bitcoin, and approximately 84 million coins for Litecoin. For these (and many other) Crypto currencies the new coins present in the economy are introduced through a process called mining, which involves solving an increasingly computationally challenging set of cryptographic mathematical problems. It is essentially a race between network nodes, as the node that solves the problem first is rewarded with a certain number of coins. Any member of the network can become a 'miner' of new coins by mining what are known as blocks.
4. A public ledger is created and maintained by all members of the network. The ledger contains an electronic record of all transactions undertaken by each bitcoin. The public ledger is formed from mined 'blocks' which are the solution to cryptographic problems that are solved to either create 'mine' new coins or to 'verify' a bitcoin transaction in the network. The ledger is known as a 'blockchain', a sequence of transaction blocks in which one can find the history of every coin from the day it was mined or created. This blockchain is periodically updated with the latest transaction block upon creation.
5. The attribution of the effort in mining a block before any other member is able, as referenced by the time stamp associated with attaining a valid solution to the cryptographic problem, is known as proof-of-work. This proof-of-work concept is also a critical tool to determination of representation in majority decision making. The Crypto currencies typically have a network consensus or majority rule component in their protocol for deciding certain key attributes of the currency.
6. Money supply is controlled by an incentive strategy, where subject to certain supply and mining conditions, the difficulty of the mining cryptographic problems is reduced or increased to control the rate of release of the new currency. In general, as the complexity of the cryptographic problems to be solved increases, the computational effort and time required to mine a coin will increase in a non-linear fashion. In particular, for Bitcoins it is currently halving at the following rate 50 Bitcoins in 2009-2012, 25 Bitcoins in 2013-2016, 12.50 Bitcoins in 2017-2020, 6.25 Bitcoins in 2020-2024 etc.

## 3. Brief Overview of Virtual and Crypto Currency Regulatory Responses

In terms of electronic currency that is non-fiat and in the form of Virtual and Crypto currency structures there is beginning to be an early literature on the analysis and understanding of such currencies and how they interact in virtual and real economies. Whilst the majority of literature on Virtual and Crypto currencies has focused on the computer science and cryptographic nature of such currency designs and distribution frameworks, there is also beginning to be a growing literature from the banking, regulation and risk perspectives to better understand these currencies as they become a more prominent part of the real economy. From the economic modelling and analysis perspectives, economic modelling of Virtual and Crypto currencies is only in its infancy, early works in this direction include the discussions on virtual currency in [Guang-zhi, 2006] and the book chapter review and references therein found in [Lehdonvirta ,2012], [Lehdonvirta, 2008] and [Lehdonvirta and Lehtiniemi, 2008]. In terms of understanding regulatory and risk based aspects of Virtual and Crypto currency banking networks there have been several recent developments, see for instance [Plassaras, 2013], [Omri, 2013], [Twomey, 2013], [Mitchell, 2014], [Kostakis, 2014], [Ogunbadewa, 2014] and [Trautman, 2014].

Given the large range of issues discussed in these papers that can arise when studying such Virtual and Crypto currencies from an economic, taxation, risk or regulatory perspective, we have decided to focus our discussion on an effort to begin to understand a particular set of issues that will arise when considering a real economy in which Virtual and Crypto currencies are free to be exchanged and used analogously to fiat counterparts. In particular focusing on a component of banking regulation yet to be explored, based on Basel II/III banking accords and in particular on Operational Risk.

To understand the significance of developing a better understanding of Operational Risk management for a banking environment in which Virtual and Crypto currencies will be present, one only needs to observe the current level of developments in such components of the financial sector, see for instance the overview in the European Central Bank extended statement [European Central Bank, 2012]. Consequently, there has been a growing awareness of the need to develop an understanding of Virtual and Crypto currencies in order to guide the understanding of regulation for such new currencies, see recent discussions particularly focussed on regulatory issues in [Lei, 2007] ,[Terando et al, 2007], [Verme et al, 2013], [Plassaras, 2013], [Teigland et al, 2013] and [Middlebrook and Hughes, 2014]. In addition, there are also a range of discussions beginning to arise that are partially opposed to regulation of Virtual or Crypto currencies see for instance [Kaplanov, 2012] and references therein.

In [Halpin and Roksana, 2009] they provide a detailed overview of recent developments in electronic money regulation. From the specific perspective of Virtual and Crypto currency there is a detailed chronological map of each country's central bank and regulatory responses[11], from which one can observe that according to the documented responses of each country, there is a wide range of dynamically evolving views on Virtual and Crypto currencies. It is clear from these regulatory responses that there is a variety of differences in opinions by central banks and regulators in how best to treat such financial instruments as they begin to interact in the real economies of different countries at a non-trivial volume.

As noted in [Plassaras, 2014], for standard fiat currencies, the International Monetary Fund (IMF) acts as an international institution whose principle task, among others, involves the co-ordination of international foreign currency exchange, see for instance discussions in [Horsefield and Garritsen De Vries, 1969] and [Simmons, 2000]. In particular the IMF is tasked with setting the regulatory guidelines for minimum standards that member nations should adhere to when designing monetary policy, so that a global financial stability is maintained. Since Virtual and Crypto currencies are formally not under the particular jurisdiction of a particular country, in particular they are not backed by any particular countries government, then there is no current reason for them to be subject to IMF guidelines. Consequently, Virtual and Crypto currencies not properly regulated have the potential to cause instability in financial currency exchange markets, such instabilities and speculative attack on fiat currencies are discussed in [Plassaras, 2014].

The most comprehensive regulatory or central bank response presented so far on the issue of Virtual and Crypto currencies interaction in the real economy comes from a report released recently by the European Central Bank, see [European Central Bank, 2012]. This is a detailed white paper discussing monetary policy implications for Virtual currencies and their potential impact on banking regulations in the EU region. This report and subsequent updates from the ECB have basically served to highlight important high-level issues such as warnings to consumers on certain key risks associated with Virtual and Crypto currencies that are currently under regulated and poorly understood from a risk management perspective. For instance, they highlight the fact that except for a few special cases of central bank Virtual currencies, such virtual currencies are more often than not failing to be contractually backed by tangible assets in a manner that is used to derive value for standard fiat currencies, nor are they typically covered by a level of legal legislation relating to tender laws in real economies or treated exactly like a tangible commodity. It is important therefore to understand better how such Virtual and Crypto currencies may derive their value without such a backing, such questions have begun to be considered and we refer the interested reader to discussions in [Irwin et al, 2005], [Lehdonvirta, 2005], [Guang-zhi, 2006], [Lin, 2008], [Lehdonvirta, 2009], [Castronova et al, 2009] and book length discussion and references therein found in [Lehdonvirta, 2014].

---

[11] http://www.perkinscoie.com/virtual-currencies-international-actions-and-regulations/

In addition to the concerns being raised regarding the ability of such Virtual and Crypto currencies to suitably maintain worth, there are also other concerns being discussed related to the need to publically educate and highlight risks that may arise unique to Virtual and Crypto currencies. They primarily relate to cyber security and e-wallet providers, since standard regulatory controls such as EU refund rights do not offer any level of protection currently for Virtual and Crypto currencies. In addition, other issues such as money laundering and taxation regulation are yet to be resolved adequately.

Other views on Virtual and Crypto currencies have been developed by different countries, for instance in France there has been a mixed response over the last few years. In 2012 in France, it was reported that a Virtual and Crypto currency exchange specifically for Bitcoins, a type of virtual currency, was established and approved by French regulatory authorities to be operating loosely speaking as a "bank", or more specifically a payment services provider under French law, see [Lee, 2012]. This decision was significant as it meant that the French authorities at the time were considering that such Euro-denominated funds would be insured by the Garantie des dépôts, the French equivalent to the US FDIC. In addition it meant that users accounts could be merged with the French banking system. Since this time additional responses have been considered, where the French Ministry of Economy declared that revenues from Virtual currency transactions in the real economy would be treated as taxable income. In addition the French Banking Federation raised concerns and is considering anti-money-laudering regulations for such currencies. Such regulatory responses continue to evolve in response to growing demand for Virtual and Crypto currencies and further support our call for greater understanding of associated banking regulatory responses and understanding such as those discussed in this paper for Operational Risk settings.

Not all EU members have been as accepting of Virtual and Crypto currencies as France, for instance in Germany in late 2011, the German Financial Supervisory Authority declared Virtual currencies such as Bitcoin (BTC) to not be considered as electronic money (e-money). This view has been evolving since this time and the current perspective of BaFin on the status of BTC's is reported in [Jens Münzer, BaFin, 2014]. Officially the response of BaFin involves the following statement:

"*BaFin has qualified BTC with legally binding effect as financial instruments in the form of units of account pursuant to section 1 (11) sentence 1 of the German Banking Act (*Kreditwesengesetz – KWG). These are units similar to foreign currencies and not of legal tender. They include value units having the function of private means of payment in barter *transactions, as well as any other substitute currency used by virtue of private-law agreements as a means of payment in multilateral settlement accounts. This makes a central issuer obsolete.*"

In addition they go on to state that one should not consider BTC as an e-money in Germany:

"*BTC are not e-money within the meaning of the German Payment Services Supervision Act (*Zahlungsdiensteaufsichtsgesetz – ZAG) because there is no issuer establishing claims against *himself by* *issuing BTC. This is different for digital currencies, which are based on a central agent (e.g. Liberty Reserve). BTC are not legal tender either, and therefore qualify neither as foreign currency nor as foreign banknotes and coins."*

Away from the EU region in large financial centers such as in countries like Singapore, there is also an active consideration of how best to handle the growth of BTC and other Virtual currencies. currently in Singapore the Monetary Authority of Singapore is considering how to tackle issues related specifically to such financial instruments pertaining to adequate verification of customer identities and reporting protocols and requirements for suspicious behaviour. Many of these types of issues will have consequences for Operational Risk management as will be discussed in future sections.

Whilst in other large financial jurisdictions such as China and Russia, they have to different degrees developed recently an outright ban at present on Virtual currencies and are aiming to prevent banks from allowing Virtual and Crypto currencies to interact in their domestic economies. For instance the Central Bank of the Russian Federation announced that[12]

*"According to Article 27 of the Federal Law "On the Central Bank of the Russian Federation (Bank of Russia)" release on the territory of the Russian Federation of surrogates is prohibited. […] The Bank of Russia has warned that Russian legal entities providing services for the exchange of "virtual currency" in rubles and foreign currency, as well as for goods (works, services) will be considered as a potential involvement in the implementation of suspicious transactions in accordance with the legislation on counteraction to legalization (laundering) proceeds of crime and financing of terrorism."*

The range of views expressed by central banks and regulatory authorities reflects the degree of uncertainty around such Virtual and Crypto currencies. It is therefore our prerogative to initiate important discussions relating to the risk management aspects that ought to be considered as Virtual and Crypto currencies continue to develop in the real economy.

Understanding the risks involved with Virtual and Crypto currency, and how these risks can be assessed, managed and mitigated so that they do not adversely impact the real economy is an essential prerequisite to good regulation. This paper describes the essential features and modes of operations of crypto currencies to highlight the drivers of Operational Risk before suggesting ways to address them, as a first step for relevant and appropriate regulation for crypto currency, to allow for this new payment system to be accepted by the wider financial community.

## 4. Types and Features of Virtual and Crypto Currencies

### 4.1 A Taxonomy of Features of Virtual and Crypto Currencies for Risk Categorisation

In the following section we outline a comparison and provide a taxonomy that highlights differences between Virtual and Crypto currencies that will be of relevance to understand when developing an analysis of the Operational Risk profiles such currencies will present if they are admitted into the banking sector en masse.

The classification of Virtual currencies and their subset Crypto currencies is non-trivial as they are multifaceted in their attributes and interactions in the real economy. The ECB report on Virtual currencies [European Central Bank, 2012] decided to classify then based on their interaction with fiat money and the real economy, they identify three possibilities: monetary flow of Virtual currencies in the economy purely by exchange mechanism with fiat currencies in order to obtain purchasing power in the real economy; direct interaction in the real economy with the ability to purchase physical goods in the real economy; or a mix of both of these attributes. However, the rules of exchangeability may evolve over time and they do not say much about the intrinsic features of the various virtual currencies that also distinguish them and importantly their potential Operational Risks if used in a banking network. Therefore, in the following sub-sections we further elaborate on key features that can be used to distinguish different types of Virtual currencies in particular the sub-class of Crypto currencies that has the largest potential for uptake in real economies.

---

[12] see Google translation provided in article: http://techcrunch.com/2014/02/07/russia-bans-Bitcoin/

In order to better understand the different forms of Virtual currencies we propose to first extend the ECB classification of them to partition the types of Virtual currencies according to a primary attribute that is whether they have a central repository and a single administrator or alternatively they follow a decentralised network consensus type administration. Such a partition of Virtual currencies is also advised in the guidance of FinCEN[13]i. We thus have two main classes of Virtual currencies, centralised Virtual currencies, and decentralised Virtual currencies. We will also refer to the latter category as crypto currencies (Crypto currencies), as the operation of these currencies is usually based on cryptographic proof provided by a network, rather than the existence of a trusted third part to verify transactions. We will now delineate the differences between these two categories of Virtual currencies, but we should note that due to the large number of Virtual currencies currently in existence, the categorisation may not apply fully to certain Virtual currencies. Note: throughout the next subsection we will first present the detail on the Crypto currencies (decentralised and cryptographic structures) and then contrast this with the same feature in the Virtual currencies which are not decentralised, allowing for a clearer understanding of the key features that can be utilised to further classify or create a taxonomy of Virtual currencies that are currently in circulation.

### 4.1.1 Specification: Centralised or Decentralised

The central bank of Canada[14] defines the notion of decentralised currency versus centralised currency as:

*"Decentralized e-money is stored and flows through a peer-to-peer computer network that directly links users, much like a chat room. No single user controls the network."*

There are numerous papers discussing the different attributes of how to design and construct protocols and architectures for both centralised and decentralised currencies, see discussions in [Garcia and Jaap-Henk, 2005] and references therein. Below we itemize the difference between Virtual and Crypto currencies:

- In Crypto currencies there is no centralised authority to exert control over any aspect of the currency, from its specification through to monetary policy. The specification of a Crypto currency is agreed by consensus by a network, and users can 'vote' for changes according to the computational power they contribute to the network.
- The specification of centralised Virtual currencies is dictated by the company that operates them (for example, Linden Labs controls the currency in the Second Life game, the Linden dollars).

### 4.1.2 Raison d'etre

The existence of electronic money has been discussed in length in Section 1, Section 2 and Section 3, here we continue the contrast between Crypto currencies and Virtual currencies with a brief comment on the motivation for their continued development:

- Crypto currencies were introduced for use in the real economy, in order to remove the need for financial intermediaries and central banking authorities, and reduce transaction costs.
- Centralised Virtual currencies were introduced for use within the environment of the authority that released the currency. Examples uses include purchases of in-game assets in massively multiplayer online (MMO) games (WoW, Second Life [Guo and Barnes, 2009]), purchases of applications (Amazon), purchases of games (Microsoft Xbox Live points until August 2013, Facebook credits until 2012).

---

[13] http://www.fincen.gov/statutes_regs/guidance/html/FIN-2013-G001.html
[14] http://www.bankofcanada.ca/wp-content/uploads/2014/04/Decentralize-E-Money.pdf

### 4.1.3 Issuance and Generation of Virtual Currencies

The generation procedure is one of the biggest factors differentiating Crypto currencies from the rest of the Virtual currencies. There are a range of academic articles discussing the different generation and money supply mechanisms for Virtual currencies and Crypto currencies, see for instance the overview of [Yamaguchi, 2004] and discussions in [Irwin et al, 2005], [Flood and Rose, 1995] and [Kroll et al, 2013]. In following the separation of Crypto currencies and Virtual currencies developed above, we note that the following basic distinctions:

- In many of the most important Crypto currencies (including Bitcoin and Litecoin), the currency is released gradually into the economy, through a process called 'mining', which involves solving a complex mathematical problem. Some currencies also feature some amount of 'pre-mining' (e.g. Auroracoin), in order to make some amount of the currency available as soon as the Crypto currency is released to the public.
- In centralised Virtual currencies, issuance is completely controlled by the central authority. On Amazon, the Virtual currency is generated instantaneously, when a user buys Amazon coins to spend on apps. This is similar to many MMO games, where some amount of Virtual currency is given to the user upon opening an account.

### 4.1.4 Monetary Policy

The money supply control and the creation process of new currency is significantly different when looking at most Virtual currencies versus Crypto currencies.

- Under Crypto currencies the money supply (and the issuance schedule for the currency) is fixed for the most important crypto currencies (Bitcoin, Litecoin). For Bitcoin, the limit is 21 million coins, the last of which will be issued (mined) over the next 100 years. The fixed money supply creates deflationary pressures.
- In centralised Virtual currencies, the central authority can exert control over monetary policy by introducing methods to control inflation, e.g. through in-game money 'sinks'. In the case of MMO games, for example, there are functions, which, once performed, give the user a benefit, but the virtual currency spent on the function is permanently removed from the game's economy [Papagiannidis, 2008].

### 4.1.5 Administration of Currency Balances

There is also in important distinction between Crypto currencies and Virtual currencies when it comes to the maintenance and monitoring of the current transactions of the currency for purchase and barter in the real economy.

- In Crypto currencies, transactions are broadcast to the network, and once confirmed, they cannot be modified or reversed. The Crypto currency balance associated with an address can therefore not be recovered if it is stolen, lost or mishandled.
- In centralised Virtual currencies the authority (e.g. Amazon, Linden Labs, Microsoft can step in to rectify an error in a transaction, reverse transactions or issue refunds as it sees appropriate.

### 4.1.6 Storage and Flows of Crypto and Virtual Currency Between Individuals

When it comes to storage of Crypto currencies and Virtual currencies and the transactions between individuals in the network, there are generally some important distinctions to note.

- With Crypto currencies, users have one or more 'wallets' storing Crypto currency addresses, each associated with a Crypto currency balance. A Crypto currency amount can be transferred from one individual to another using a cryptographically signed transaction, which has to be broadcast to the network and confirmed. Besides interacting with one another directly, users can also trade Crypto currency with one another through a number of exchanges.
- Centralised Virtual currency balances are stored within the environment in which they were issued. Users can use Virtual currency amounts to purchase virtual goods from the platform/website, or from other users, but cannot generally transfer such currency outside this online environment. Exceptions to this can occur when users act outside the intended rules, for example, by transferring the details to an entire account (and therefore the contents of the account, which could contain Virtual currency, also) [Lehdonvirta, 2014].

### 4.1.7 Exchange Rates and Flows between Crypto/Virtual Currencies and Fiat

The ability to exchange Crypto currencies and Virtual currencies with fiat currency and the rate of exchange is now provided in several denominations such as Crypto currencies versus USD or Crypto currencies versus EURO etc. on a number of international electronic exchanges, such as Coinbase.com; Cryptsy.com; Bter.com; and Coins-e.com[15]. There are some important distinctions to be noted generally between Crypto currencies and Virtual currencies when it comes to exchange rates. Recent studies have begun to analyse these from a modelling perspective and from a monetary policy and regulatory perspective, see [Plassaras, 2014].

- In Crypto currencies, flows are bidirectional. Exchange rate varies, according to supply and demand (Bitcoin, Litecoin are traded on exchanges with sufficient volumes for the price to reflect supply and demand, and exchange rate moves are not due to just illiquidity jumps)
- In centralised Virtual currencies, flows can be single-directional (e.g. Amazon coin, Zynga in-game currencies, Microsoft Xbox Live Points, Facebook credits) or bi-directional (Linden dollars in Second life, Diablo III markets). The exchange rate is usually fixed (e.g. Amazon coin), where in this case the retailer also offers discounts for the purchase of higher quantities of coins. However, in the case of centralised Virtual currencies that feature a currency exchange, the rate can also be controlled by central authority (e.g. Linden Labs controlling Linder dollar to USD exchange rate [Papagiannidis, 2008]).

### 4.1.8 Value Generation

The economic understanding of the value generation mechanism for many Virtual currencies and Crypto currencies is still at early stages, see for instance the book length discussion in [Lehdonvirta and Castronova, 2014] and other discussions in [Malaby, 2006], [Castronova et al, 2009], [Rogojanu and Badea, 2014] and the recent overview in [Plassaras, 2014].

- There are two sources of value for a Crypto currency:
  - Due to the size of the network using it, as a Crypto currency can therefore derive some of its value from network effects [Courtois et.al., 2013]. As the relationship between network value and size is superlinear [Briscoe et.al., 2006], the implication is that the value of the most popular Crypto currency (Bitcoin) is much higher than that of other Crypto currencies with fewer users, and this is reflected in the 'market capitalisation', or the total value of coins in existence.
  - Due to the cost of producing (mining) an amount of the Crypto currency. As the network grows, the computational power required to produce a coin increases, and this should increase the value of the Crypto currency also.

---

[15] A comprehensive list of such currency exchanges currently available: http://planetbtc.com/complete-list-of-Bitcoin-exchanges/

- There are other value generation mechanisms for centralised Virtual currencies related to the interplay between virtual economies and real economies, which has been studied in detail in economic analysis such as [Plassaras, 2014] and [Lehdonvirta, 2014]. Virtual economies are observed in, for instance, MMO and in life simulation games which may have taken the most radical steps toward linking a virtual economy with the real world. This can be seen, for example, in Second Life's recognition of intellectual property rights for assets created "in-world" by subscribers.

#### 4.1.9 Currency Area

The currency area is the geographic area in which a given currency is accepted as a means of payment.

- There are no limitations to where Crypto currencies can be used in the real economy, but so far only a limited number of retailers are willing to accept payment in Crypto currencies. The majority of transaction activity is for speculation, in crypto currency exchanges.
- In centralised Virtual currencies, the currency area only comprises of the website (or subset of the website) which operates the virtual currency. For example, Amazon coins can only be used for the purchase of apps from the Amazon app store. For MMO games, this is the virtual world of the game.

## 5. Operational Risk in Crypto Currencies and Possible Mitigation

### 5.1 Operational Risk Vulnerabilities and Exposures of Crypto Currencies

Due to the differences specified in the previous sections between Virtual currencies and Crypto currencies, we can only see the possibility of Crypto currencies arising as an alternative or complement to fiat currency for transacting in the real economy en masse. While it will be possible under Virtual currencies to use the proceeds of activities in some MMO game to make real-world profits, the effect on the real economy will be tangential, at best. In addition, we argue that the drivers for Operational Risk's under Virtual currencies versus Crypto currencies are of a substantially different nature and intensity. We believe that Operational Risk will be both more numerous and more intense for Crypto currencies. These currencies are also those, we believe, with the biggest potential for widespread acceptance and development in the future. We therefore restrict the scope of this section to the analysis of operational risk for Crypto currencies. We will now focus on the operational risk drivers for financial institutions transacting in Crypto currencies.

In order to clarify the issues at stake in the management, and also in the regulation of virtual currencies, we propose a structure of risk identification starting with an identification of risk drivers that allows one to formulate risk mitigation actions addressing the causes of the risks rather than the symptoms, for an effective Operational Risk management response. Next, we map the risks generated by these drivers to the Basel categories, to define the specific risk profiles of Crypto currencies in relation with the Basel regulatory guidance. Table 1 summarises this mapping, down to the level 2 definitions of Operational Risk types and provides some brief examples of these risk types under the Basel II/III regulatory guidelines one can expect to arise in a banking environment directly of relevance to Operational Risk for Crypto currencies should they be introduced to banking networks and increase circulation in the real economy.

**Table 1 :** Basel Categories of relevance to Operational Risk for Virtual currencies and Crypto currencies, [BCBS (2006, pp. 305–307)].

| Risk category level 1 | Risk category level 2 |
|---|---|
| 1. Internal fraud | 1.1 Unauthorised activity<br>1.2 Theft and fraud |
| Examples include:<br>- *misappropriation of assets* (i.e. Crypto currencies through for instance theft of private and public keys); and | |
| | - *tax evasion* (this issues has already raised concerns with several regulatory authorities). |
| 2. External fraud | 2.1 Theft and fraud<br>2.2 Systems abuse |

Examples include:
- *theft of information* (this may include virtual wallet addresses, public and private keys as well as other personal identifications such as transactions made between members of the Virtual currency and Crypto currency networks);
- *hacking damage* (permanent corruption or destruction that is irreversible for portions of the currency or members accounts); and
- *third-party theft and forgery* (theft of virtual currency from exchanges, virtual wallets and storage facilities, and eventually banks and depository taking institutions should they accept Virtual currencies and Crypto currencies in the future).

| | |
|---|---|
| 3. Employment practices and workplace safety | 3.1 Employee relations<br>3.2 Health and safety<br>3.3 Diversity and discrimination |

Not perceived at this stage to be particularly relevant for Operational Risk primarily from the perspective of Virtual currencies and Crypto currencies.

| | |
|---|---|
| 4. Clients, products and business practices | 4.1 Conduct<br>4.2 Advisory activities and mis-selling<br>4.3 Product Flaws<br>4.4 Improper Business or market practices<br>4.5 Customer or client selection and exposure |

Examples include:
- *market manipulation* (currently most electronic exchanges for Virtual currencies and Crypto currencies are under regulated if they are regulated at all);
- *antitrust* (there is potential moral hazard associated with privately or consortium based Virtual currencies);
- *improper trade*;
- *product defects* (there may be unknown design problems with security, coin generation, verification etc. yet to be discovered in Crypto currencies protocols that could be exploited by malicious users);
- *fiduciary breaches* (since fiduciary relationships are by their very nature relationships of good faith, they may involve a variety of obligations depending on the exact circumstances. In the context of Virtual currencies and Crypto currencies dealings, since there is no central authority monitoring these currencies, they are open to manipulations which may also impinge on fiduciary expectations typically required of corporate directors, officers and risk management functions of financial institutions. In addition, the secrecy and lack of monitoring of legal ownership of Crypto currencies and their attribution to particular persons, may allow for conflict of interest and lack of fair practice with regard to the best interest of the employer/principal, free of any self-dealing, conflicts of interest, or other abuse of the principal for personal advantage.
- *account churning* (special forms of account churning may arise in Virtual currencies and Crypto currencies which are related to the incentive commissions paid for proof-of-work as well as incentives paid by transaction fees verification. For instance in some Crypto currencies when the output value of a transaction is less than its input value, the difference is a transaction fee that is added to the incentive value of the block containing the transaction.)

| | |
|---|---|
| 5. Damage to physical assets | 5.1 Disasters and other events |

Examples include:
- terrorism (cyber terrorism and attacks on networks and storage facilities may be initiated to benefit certain members or destroy Crypto currency account details and digital records);

| | |
|---|---|
| - vandalism (cyber vandalism from hackers) | |
| 6. Business disruption and system failures | 6.1 Systems |
| Examples include:<br>- Software failures (software specific to a particular Virtual currency or Crypto currency may get upgraded or modified resulting in failures in mining, transaction verification or even encryption);<br>- Hardware failures (core nodes on the mining network required to process block chains and perform network verifications may fail resulting in extensive delays in transaction processing).; | |
| 7. Execution, delivery and process management | 7.1 Documentation; Transaction; A/c Management; Reporting; Distributor; Supplier |
| Examples include:<br>- data entry errors; accounting errors; failed mandatory reporting; and negligent loss of client assets (all these may affect different aspects of Crypto currencies for instance with regard to storage of virtual wallets, private and public encryption keys). | |

In Table 2, we provide a summary of what we consider are a collection of important core vulnerabilities and exposures under Operational Risk for Virtual currencies and Crypto currencies within a banking environment. The categorisation of drivers into exposures and vulnerabilities, allow also the selection of appropriate risk mitigating actions. In the following subsections we discuss in more detail the vulnerabilities and then the exposures outlined below. We consider in this paper that vulnerabilities: *are weaknesses, failing processes or inadequate features*. The risk mitigation of vulnerability drivers must include the reinforcement of weaknesses, solutions to problems, maintenance and controls, we discuss examples of mitigating actions in Table 3.

**Table 2 :** Summary of Vulnerabilities and Exposures of Crypto currencies to Operational Risk.

| Vulnerabilities | Exposure |
|---|---|
| 1. Decentralised governance | 1. Multiplicity of jurisdictions |
| 2. Peer to peer verification | 2. Multiplicity of micropayments |
| 3. Transaction irreversibility | 3. Hardware reliance |
| 4. Anonymity | 4. Software reliance |
| 5. Handling of sensitive information (private keys and virtual wallets) | |
| 6. Price or value instability | |
| 7. International regulatory risk | |

From the identification of these risk drivers for Crypto currencies, such as Bitcoins, one can now map these to the Basel II/III categories as presented in Table 3.

**Table 3 :** Risk drivers of Crypto currencies such as Bitcoins mapped to Basel II/III categories

| Risk Drivers | Operational Risks | OR Basel Categories | Mitigating actions (non exhaustive list) |
|---|---|---|---|
| **Vulnerabilities** | | | |
| Transactions Anonymity | - Money Laundering<br>- Fraud<br>- Legal risk<br>- Compliance | 2.1.<br>4.4. (AML, KYC, Regulatory breach and non compliance) | Forcing identification, at minimum via a registered IP address; tax file identification when registering e-wallet |
| Absence of legal | - Governance risk | 2.1. | If in Virtual currency |

| tender | - Abuse of power<br>- Price manipulation<br>- Fraud<br>- Price volatility<br>- Changes of protocols | 1.1.<br>4.4.<br>6.1.<br>7.1.<br>7.2.<br>7.3.<br>7.4. | and Crypto currency deposit taking banks: codify network and governance powers |
|---|---|---|---|
| Peer to peer verifications of transactions | - Double spending attacks | 2.2.<br>6.1. | High computational power |
| Transaction irreversibility | - Aggravated losses in case of errors on theft on private keys (cfr. next driver) | 1.1.<br>2.1.<br>7.3. | |
| Private key as unique mean of access | Loss of value to due:<br>- External fraud<br>- Internal frauds<br>- Processing errors<br>- Damage to physical assets<br>- Attacks | 2.1.<br>2.2.<br>1.1.<br>6.1.<br>7.2.<br>7.3. | Similar risk mitigation techniques applicable to sensitive information in traditional banking |
| Publication time gap | - Fraud in double spending<br>- Fake transactions,<br>- Attacks,<br>- Transaction malleability | 2.2.<br>2.1. | |
| **Exposures** | | | |
| Multiple jurisdictions | Regulatory breach<br>Tax avoidance<br>Tax law compliance | | |
| Micropayments | Internal fraud<br>External fraud<br>Systems failure<br>Process errors<br>Reporting | 1.1.<br>2.1.<br>6.1.<br>7.4<br>7.3.<br>7.5. | |
| Hardware Reliance | Exposure to hardware failures | 6.1. | |
| Software Reliance | Exposure to software failure<br>Drivers of impact of data theft | 2.1.<br>6.1. | |

## 5.2 Operational Risk Vulnerabilities of Crypto Currencies

As noted in the introduction, the dominant crypto currency thus far is Bitcoin, with a market capitalisation of approximately $6.25 billion. Apart from Bitcoin, Litecoin, Nxt and Dogecoin have seen steady pickup in volume as well market value, but still remain very far from Bitcoin in terms of valuation, and the total market capitalisation of all other crypto currencies is less than $2 billion[16]. For this reason, we will focus the remainder of discussion in this paper on the particular Crypto currency known as Bitcoin. Therefore, to proceed it will be useful to highlight a few important features of Bitcoin to understand better vulnerabilities and exposures such a Crypto currency will create from an Operational Risk perspective, should it be allowed into the banking sector.

---

[16] http://coinmarketcap.com/

To proceed with analysis of vulnerabilities and exposures, specifically for Crypto currencies like Bitcoin, it will aid discussion to understand the basic feature of public-key encryption adopted by many Crypto currencies like Bitcoin. In particular, Bitcoin relies on public-key cryptography, an asymmetric key encryption scheme which is used for encrypting messages and verifying the originator of a message. A user wishing to communicate, using a public-key cryptography scheme would have two keys: A public key that is available for everybody to access, as its name implies, and a private key that must be kept secret. As an example, consider a user (B) that wants to send a message to user A in a public key cryptography scheme. Then B would have to obtain A's public key, obtains the `ciphertext' (or the encryption transformation of the message defined by the public key), and sends it to A. A can then decrypt this message using her private key. The process is described in detail in [Vanstone et. al., 1996].

In addition, we also note that Bitcoin uses the concept of digital signatures to ensure non-repudiation: that is, a third party can easily verify whether a particular signatory has signed a message, using only information that is publicly available (the signatories private key). The process of sending a payment therefore proceeds as follows:

1. A obtains B's Bitcoin `address', a string of alphanumeric characters which represent a destination for a Bitcoin payment.
2. She creates a message with the address and number of Bitcoins she wants to send to B, and signs it with her private key.
3. She then broadcasts this message to the network, and the network nodes are able to verify that the message originated from A, using her public key.

It is important, then that the private key remains secret. If another individual (say C) has access to the private key, they would be able to create a transaction message and sign it as if they were A, possibly transferring units of the currency to their own address, and the Bitcoin balance of A's address will be compromised.

Once a transaction occurs, it is broadcast to the network, and is verified by the network nodes. It is then inserted into the 'blockchain', a shared public ledger of transactions. It is therefore straightforward to identify the last owner of a particular unit of currency, further technical details on transaction protocols for Crypto currencies like Bitcoin are discussed in [Andrychowicz et. al., 2013a] and the references therein.

### 5.2.1 Decentralised Governance

As noted in the background on Crypto currencies provided above, they operate via a peer to peer network, independent of a central authority or central banks, this feature though appealing from a conceptual standpoint for many advocates of Crypto currencies like Bitcoin, also turns out to be an Operational Risk vulnerability as will be detailed in this subsection.

The decentralisation of Crypto currencies means that all functions such as issue, transaction processing and verification are managed collectively by this network.

In his seminal paper on the most popular Crypto currency, the Bitcoin, Nakamoto (2008) states:

"*The system is secure as long as honest nodes collectively control more PCU power than any cooperating group of attackers nodes.*"

This is due to the peer-to-peer review system of validation of transaction, where validating power comes with CPU power, in a system similar to "*one-CPU-one-vote*". Indeed, the confirmation (of not double spending) of transaction requires the knowledge of all previous transactions and their times in order to decide what comes first.

Nakamoto argues: "*As long as a majority of CPU power is controlled by nodes that are not cooperating to attack the network, they will generate the longest chain and outpace attackers*".

The weakness of this argument is, in our view, the assumption that financial crime or other organised consortiums would not attempt to outpace genuine network controllers. There is, arguably, a non null possibility of systemic failure of a Crypto currency network such as Bitcoin in case of a co-ordinated attack. Systemic risks are typically to be addresses by national and international regulators, possibly by ensuring that extreme CPU power is allocated to recognised network controllers such as national agencies and agreed private parties.

More likely than co-ordinated attacks are the other failures that may arise under due to the possibility of an alteration of protocols the dictate the processing and creation of Crypto currency such as Bitcoin. In the case of Crypto currencies like Bitcoin, any aspect of the protocol can be altered by consensus. currencies that employ the proof-of-work system are vulnerable to miners (or groups of miners) that accumulate large computational resources for mining and verification. In such a currency, a pool that controls more than the majority of the computational power can impose conditions on the rest of the network, and engage in malicious activities, see discussion in [Kroll et. al., 2013] and [Decker and Wattenhofer, 2013] and references therein for more specifics on the technical details.

In the case of Bitcoin, currently two mining pools control more than 50% of the computational power of the network, (see Figure 1) consensus appears to already be within grasp. Changes to the protocol include even the most fundamental part of the specification, such as the coin-introduction function: The Bitcoin protocol specifies that there is fixed total number of Bitcoins, and if this were to change, it would cause major changes both to the value of the currency and perhaps to the network itself, as many might ignore such a fundamental change. Interestingly, analogies can be drawn between governance issues for the control of crypto-currencies network and corporate governance issues for corporate control, see discussion on this aspect in [Chapelle and Szafarz, 2005].

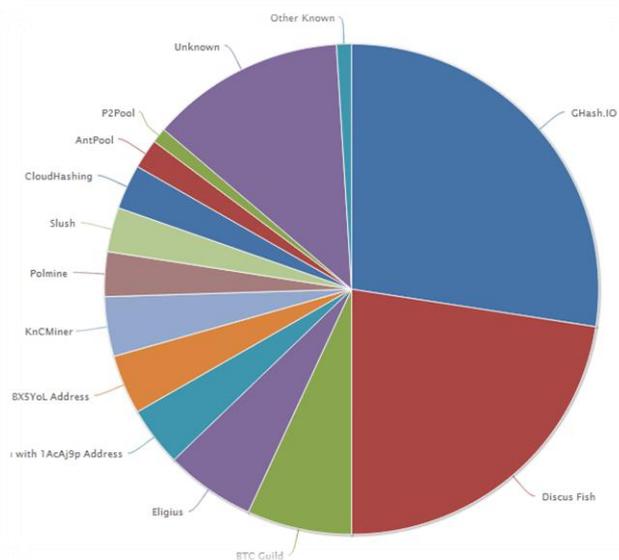

**Figure 1 The computational (mining) power of the various mining pools on the Bitcoin network as of 17/08/2014. GHash.IO controls approximately 28%, while Discus Fish controls approximately 23% of the network mining power. Source: https://blockchain.info/pools**

Besides the absence of a legal tendering authority, the decentralised, network based management of virtual and of crypto currencies in particular acts both as a risk driver and as risk mitigation. On the risk side: the fight for network control, politics and power struggles and possible instability of protocols. On the mitigation side: there is no single point of failure. In contrast with traditional payment systems, as there is no central intermediary; any node may drop out of the network and new nodes can enter at any time, and are compensated according to the computational power that they contribute. In theory, this should lead to a more resilient network, while also protecting against bad actors trying to affect change in the Bitcoin protocol, as changes have to be approved by the majority of the computing power in the network.

As the Bitcoin system has evolved, however, both the mining power and the decision making authority of the network are increasingly being monopolised by a limited number of mining pools. These are collections of network nodes that have decided to pool their computational power, in order to ensure a more consistent payout from their mining operations. [Gervais, 2013] notes several other limits to decentralisation, which go against the original intentions of [Nakamoto, 2008].

In the context of Crypto currencies such as Bitcoin there is also a challenge for financial institutions that may consider accepting Crypto currencies, such as Bitcoin, to overcome. This challenge is related to the recently discovered issues pertaining to a potential exposure in the cryptographic protocol for such a currency which allows for a small window of time in which manipulation of the transaction details may be undertaken, either by human error or under fraudulent activity. This issue was described in [Andrychowicz et. al., 2013b] and is known as **transaction malleability**. This was postulated to have been one of the contributing factors to the large losses incurred at Bitcoin exchange Mt Gox, although this has been disputed recently in [Decker and Wattenhofer, 2014].

From an Operational Risk management perspective this transaction malleability manifests as weakness that enables the construction of functionally equivalent signed messages of transactions, which may then be submitted to the network. Under certain circumstances, described in [Andrychowicz et. al., 2013b], this will cause one of the original parties to the transaction to lose money. Alternatively, it could also be a way to trick an exchange into believing that a transaction had failed, when in fact an equivalent transaction had been confirmed by the network, see discussion in [Decker and Wattenhofer, 2014].

Another consequence of fraudulent activity that may arise due to decentralise governance may include market manipulation by the Crypto currency worth with respect to the exchange rate of the Crypto currency with other Virtual currencies and fiat currency. This could be manifest within and localised within a particular banking institution or it may arise amongst many institutions with a group of malicious agents seeking to profit jointly from fraudulent market manipulations. This may arise since presently the exchanges upon which the Virtual currencies are traded and exchanged are unregulated. Unless there is a regulatory oversight of these exchanges then there is an exposure present that may allow banking institutions admitting Virtual and Crypto currencies to have individuals that may partake in the manipulation of the exchange rates of Virtual currencies to their commercial advantage. This may include price fixing by members of the bank, which if investigated by regulators and financial authorities in a given jurisdiction may expose the affects financial institutions and Banks to regulatory fines and litigation which would contain clear components directly under the remit of Operational Risk.

### 5.2.2 Peer-to-Peer Verifications of Transactions

We will explain in this subsection how the Peer-to-Peer and proof-of-work aspects of Crypto currencies like Bitcoin can also create several vulnerabilities, which if exploited have the potential to generate substantial losses that may be attributed to Operational Risk loss categories.

The peer-to-peer verification process in Bitcoins is an example of a `proof-of-work' system discussed in the section on Crypto currencies, in which the effort required to obtain a reward is great, while the effort required in verifying the validity of the work done is small. For example, Bitcoin uses the Hashcash proof of work system[17]. In this system, the miner is required to perform a partial inversion of a hash function. It should be noted, that Bitcoin is one of the faster protocol designs with regard to transaction verification. Other Crytpo currencies such as Litecoin use alternative approaches such as scrypt[18], which requires more memory, and thus is more difficult to design specialised hardware for the computer task of verification.

As a consequence of peer to peer verification, there is, even in the faster protocols for Crypto currencies like Bitcoin, a delay in processing a transaction which can be in the order of 10's of minutes of gap between the execution of a transaction and its publication to the network and registration on the Bitcoin ledger blockchain. Typically, in Crypto currencies, transaction blocks are published to a network periodically, as in Bitcoin, where average block publication times are 10 minutes, with a standard deviation of 15 minutes [Karame, 2012]. Hence, the network will only be aware of a transaction after a certain interval of time.

Consequently, these delays, though seemingly innocuous, actually present as a serious vulnerability for Crypto currencies. They present a vulnerability related to the risk of fraud, system attacks, double spending and fake transactions. In these waiting periods, an adversary could attempt to uses the same Bitcoins in multiple transactions, and if the goods are released instantaneously, as in the case of paying for digital goods, like digital video and audio, this may easily lead to losses for a vendor. It seems to be a core operational weakness currently in crypto currencies and for Bitcoin in particular, see discussion in [Karame, 2012].

This vulnerability of Crypto currency is known as double spending and has been a major issue to be resolved, which we argue is still to be understood in detail from the perspective of vulnerabilities that may eventuate if exploited in significant Operational Risk losses for financial institutions accepting Bitcoin transaction processing. In some aspects it is analogous to the problem in fiat paper money of check kiting. Under a Virtual or Crypto currency offline payment system the notion of double spending arises due to a failure of digital cash schemes to prevent the possibility to spend a single digital token twice. Since, unlike physical token money such as coins, electronic files can be duplicated, and hence the act of spending a digital coin does not remove its data from the ownership of the original holder, therefore other cryptographic approaches are required to prevent such frauds.

Hence, the act of proof-of-work and in particular the delay required to verify transactions validity in Crypto currencies like Bitcoin admit the possibility of `double-spending' or the use of the digital currency in multiple transactions by the same individual. Several approaches have been proposed to overcome this problem, Typically a digital payments system would have to have an alternative method to verify that a transaction has occurred and ownership of the electronic money tokens had been transferred to another individual.

A detailed description of particular forms of double spending attacks that may circumvent these attempts to prevent such fraudulent behaviour are starting to be studied, see for instance discussions in [Demers et al, 1998], [Pointcheval and Jacques, 2000], [Nakamoto, 2008] and [Karam et. al., 2012]. It has recently been discussed in [Karam et. al., 2012] how such a double spending attack my manifest. We spare the technical detail in this brief description, but find it informative to understand better this Operational Risk vulnerability to provide a basic analysis in the case of Bitcoin.

Fraudulent transactions in the Bitcoin Crypto currency network can be generated in a double-spending attack and may cause losses, even when they have been confirmed by network nodes. If an adversary creates two payment messages (containing a genuine and a fraudulent transaction) using the same Bitcoins, but to be sent to two different parties, and broadcasts them to the network simultaneously, it is likely that different network nodes will receive the two messages in different order. They will verify the earliest message as being the valid one, and reject the second, and attempt to publish them in a transaction block. Two nodes may then publish transaction blocks with different transactions (a `fork' of the network), and depending on how this is resolved, either the genuine or the fraudulent transaction may be confirmed.

---

[17] Proof-of-work in Bitcoin: https://en.Bitcoin.it/wiki/Proof_of_work
[18] Verification Protocol of Crypto currency Litecoin: https://litecoin.info/Scrypt

The probability of a successful attack depends on the computational power of the attacker, compared to that of the network. [Rosenfeld, 2014] analyses the stochastic processes underlying typical attacks and finds that if the attacker's computational power is 10% of the total network computational power, one requires 2 confirmations to keep the attack success rate below 10%, 4 to have it less than 1%, and 6 to decrease it below 0.1%. Standard practice is for a vendor to wait for n confirmations of the paying transaction, before providing the product, with 3-6 confirmations being the norm.

Another vulnerability that arises from the proof-of-work feature of Crypto currencies, that may manifest in fraudulent activities that could generate Operational Risks includes the idea of co-operative `selfish' mining strategies. This vulnerability involves the case in which a group or consortium of miners, who are critical to the verification of transactions and creation of new currency, that control at least 1/3 of the mining power of the network can mine a disproportionately high number of Bitcoins [Eyal and Sirer, 2013]. The consequence of this is that they may fashion a mechanism in which they then selectively publish transaction blocks they have discovered (mined), causing the `honest' majority to needlessly spend computational power in mining the same blocks with little reward or outcome. In addition, this vulnerability may also be turned into a form of preferential transaction verification that may open its way to fraudulent behaviours resulting in Operational Risks such as bribery for transaction processing preferences, moral hazard in processing one's own transactions with priority and account churning for transaction fees.

We note that the notion of proof-of-work is used in the most important Crypto currencies (Bitcoin, Litecoin). However, there are also other system protocols and designs that have been proposed, such as proof-of-stake [King, 2012]. These systems are also not without vulnerabilities that could result in significant Operational Risk losses. For instance, in the case of proof-of-stake systems, Crypto currencies based on this approach would be vulnerable to attacks by major shareholders.

Hence, until there is a universal consensus and understanding of these features, this aspect of Crypto currencies still remains a serious vulnerability from an Operational Risk managers perspective, when assessing the risk of allowing a financial institution to begin to accept processing of Crypto currency transactions.

### 5.2.3 Transaction Irreversibility

Another form of vulnerability that arises in Crypto currencies is the notion of transaction irreversibility. It is important to note the difference between standard electronic transaction payments and those of Virtual currencies like Bitcoin which relates to irreversibility of any payment or transaction, once it enters the block-chain [Barber, 2012]. In other words, a data entry error cannot easily be corrected. If for example, in the transfer of a large amount of Bitcoins by the bank, the amount to be transferred is mistakenly switched with the transaction fee, then the miner verifying the transaction can keep the fee and the bank has no recourse for complaint, in order to reverse or modify the transaction. This is both because of the specification of the currency, where the publication of a transaction `block' depends on all previous blocks, as well as the lack of a central authority that could generate a new transaction to offset the original one. This is one of many possible errors that may occur for Virtual currency transactions and payments. Due to this feature of Crypto currencies relating to transaction irreversibility, a bank may not be able to recover the virtual currency and will incur a financial loss. Clearly, such vulnerability will affect all retail banks with electronic payment systems.

A second type of vulnerability that could generate Operational Risk losses may also arise in the context of transaction irreversibility, for financial institutions or deposit taking institutions such as a Banks, that are considering accepting Crypto currencies. This second example of a vulnerability due to transaction irreversibility relates to the monitoring and maintenance of customer account virtual wallets. This involves the fact that the bank requires continuous access to the block chain, to ensure that it can keep track of the ownership and status of its customers' account balances, ie. number of Bitcoins. As such, there is a threat that a bank will have to halt operation, or limit its operations due to a hardware or software fault. In particular, telecommunication problems (due to internet downtime, a power outage etc) may affect the bank's ability to carry out its operations. If during such an event there are transactions processed unbeknownst to the bank, then this may result in Operational Risk losses due to the irreversible nature of Bitcoin transactions and verifications, which would be processed by the network but remain unknown to the banking institution until their own network or communication problem is resolved.

Therefore, processing large volumes of transactions and waiting for authentication of such transactions may result in alternative protocols for verification being adopted in different banking institutions. This could manifest in Operational Risk losses if for example, say retailers dispensing inexpensive items such as consumables may begin to generally accept a Bitcoin payment made from a consumers account in such a bank with relaxed or modified authentication protocols, which due to time constraints of process may choose to avoid waiting for 3-6 confirmations of verification, as is the norm for larger transactions. Then, if however, there are insufficient funds or there is some anomaly in the transactions, and the bank has already accepted the transactions due to their lax verification protocol not immediately detecting this problem, then this may then result in the trading entities in question being denied the funds. Consequently, this may result in a liability or loss from the banks capital in the form of compensation for the incorrect processing as well as legal costs for the recovery of the incorrect transaction.

Finally, there is also the serious vulnerability regarding transaction irreversibility related to cyber crime and hacker attacks on banking networks and customer virtual wallets or Crypto currency bank acocunts. If such attacks as those described in detail in [Reid and Harrigan, 2013] are not carefully avoided by banks and financial institutions accepting Bitcoin deposits, then the irreversible nature of such attacks leaves the banks capital open to significant losses resulting form compensation. Large cyber thefts of Crypto currencies are progressively reported to occur, an example in [Reid and Harrigan, 2013] they discuss the real case of a theft of 25,000 Bitcoins, worth in current exchange rates on 12/9/2014 around 11.9mil USD. The irreversible nature of such thefts leaves potential for large vulnerability in banks accepting to hold customer accounts in Crypto currency. It may also be extremely difficult to even detect such cyber crimes immediately, see example of detailed analysis why and the steps taken by the attacker to avoid identification in [Reid and Harrigan, 2013]. There have also been a number of other large high profile thefts being reported in world print news since 2011 onwards, and clearly there is a security vulnerability and subsequent capital loss due to irreversibility faced by institutions holding bitcoin insecurely that manifests as a serious source of potential Operational Risk losses.

### 5.2.4 Anonymity

Anonymity has the potential to be an important vulnerability that could result in Operational Risk losses from Crypto currencies. There are several aspects of this vulnerability that must be explored from the perspective of financial risk which include features such as transaction processing privacy of customers, money laundering implications, taxation on accounts and some aspects of these may result in Operational Risk losses. We discuss these in more detail in this section.

Currently, a Bitcoin address with its pair of public and private keys, are the only requirements to undertake transactions in Bitcoins. The Bitcoin address is not registered to a named individual; only the possession of the private key gives control over the balance associated with the address. While the complete address history of every Bitcoin is traceable, the controllers of those addresses are not necessarily easily identifiable, making the transactions anonymous.

However, total anonymity is only guaranteed under certain circumstances, i.e. employing anonymising software and transacting directly with individuals that are also similarly careful in protecting their identity. [Brito and Dourado,2013] refers to the anonymity of crypto currencies transaction as "pseudonymity", since in practice, transacting through a website will mean that an individual would, at the very least, leave a digital footprint in the form of an IP address, which one could then be tied to a physical address. The use of Bitcoin web `wallets' is an example.

One way to understand how this anonymity can lead to a vulnerabilities from an Operational Risk perspective is to consider the following two examples. In the first case, consider for instance what the implications of anonymity in processing transactions may have. Without the ability to properly identify parties on both ends of a transaction, Crypto currencies are at the mercy of money laundering attempts. For financial institutions and banks considering the acceptance of Crypto currencies they need to be careful therefore to make sure they may still satisfy regulations on AML (anti-money laundering), KYC (Know your customer) and PEP (Politically Exposed persons), just to name a few. These regulations can make the suppression of some aspects of the anonymity a necessary condition before even starting to consider the acceptance of Bitcoin transactions or other Crypto currencies. This is critical to avoid, for instance the acceptance of deposits in say Bitcoins that were obtained through fraudulent activity, either directly by the client making the deposit into the e-wallet hosted by the bank accepting the Crypto currency or indirectly, unbeknownst to the client making the deposit. See [Moser, 2013] for a review of the effectiveness of current AML tools, as well as Bitcoin services that could hinder AML efforts.

In the second example, we look from a different perspective at the consequence of anonymity and the peer-to-peer verification structure and blockchain of Crypto currencies such as Bitcoin. Under the peer-to-peer review system, one requires a full traceability of operations. Therefore, by its very protocol design, such Crypto currencies allow anyone to see the balance and the detail of every transactions operated by any address (testable on biteasy.com). This is indeed part of the verification process, and works currently due to the fact that there is anonymity in the account holders details never being required.

However, one would expect that in the future clients, when setting up an e-wallet in a financial institution or bank accepting deposits in Crypto currency, would provide a taxfile number or identification of this form so that any interest credited to their accounts in Virtual currency denomination or in fiat currency can be appropriately considered for taxation purposes. This on its own has interesting economic and legal implications relating to the so-called anonymity offered by transacting in Virtual currencies in the real economy. The anonymity would be present up to the transactions being processed from such accounts registered to a particular client of the bank. Once an account is opened it would have to be linked to taxation details of the client, therefore providing an identity to transactions in the block-chain, at least related to this portion of a transaction sequence. This can also be valuable for regulators and cyber security agencies tracking criminal activity and money laundering in such Virtual currency international exchanges.

Now, one can easily see where the vulnerability arises from an Operational Risk perspective. If however, the account owners addresses were no-longer anonymised, at least internally within a bank holding Crypto currency deposits and within a particular countries government revenue and taxation office database, due to the need to satisfy regulation on taxation and AML, then an attack or leak of such account details would present as a huge Operational Risk vulnerability to a number of potential losses from risks due to litigation, breach of privacy as a result of what may be claimed to be product defects. To understand this risk, we note that having obtained the personal identity of an account holder that could be associated to a virtual wallet or public keys IP address, then all their associated transactions immediately become public knowledge through analysis of the publically available block chain. Hence, upon theft of fraud related to these private account details, the theif gains much more than just private information, as would normally be the case with standard bank accounts, they also gain all information related to their transaction history in their Bitcoins. It would immediately allow people to see exactly who was transacting with whom, when, where and how much was transacted. Clearly, we therefore see that the resulting verification structure of Crypto currency blockchains, in the case of violation of account anonymity is not such a convenient feature for operators. In fact it is dramatically different from the principle of secrecy of the traditional banking operations in which no account transactions are ever disclosed to the public.

We note, that there may be some alternatives to circumvent some aspects of this issue, for instance Bitcoin user that may not appreciate the full disclosure of his operation can install a transaction system that will generate a different address every time he executes a payment. This makes tracking more difficult for dishonest parties, but also for the regulator and financial institution or bank who attempts to accept such Crypto currency transactions and deposits.

A third source of vulnerability, associated with Crypto currencies like Bitcoin that admit a public blockchain ledger of all transactions and account details in the transactions in terms of address, involves the issue of banks accidentally or wilfully accepting Bitcoin deposits that are proceeds of crime or have been involved in some criminal activity. The interesting question here is how best to decide what to do with such Bitcoin transactions and deposits. This is interesting since technically every transaction of every Bitcoin is logged in detail in the blockchain, providing a detailed history of all accounts the Bitcoin has transacted through. Therefore, if regulators and law agencies held a record of fraudulent activities related to particular IP, e-wallet or QR addresses then a question would arise relating to how and when should it be suitable to accept the Bitcoins into a given banks deposits. This is the Crypto currency equivalent of the issue that arise for banks accepting cash deposits where serial numbers are available to allow law enforcement agencies to track proceeds of crime or thefts. In such cases, there is generally a moratorium period after which the cash is legally allowed to be deposited in a bank or financial deposit taking institution. Hence, one of the vulnerabilities faced by banks accepting deposits in Crypto currency is how to manage the risk that a particular subset of the Bitcoins held in a bank was previously (some transactions ago) used in a criminal transaction and can thus be confiscated under the proceeds of crime act, or the equivalent law in other countries. Such considerations need to be made with banks accepting Bitcoins or they may appear to be assets which would later be removed from the bank leading to Operational Risk losses.

One possible approach to tackling this Operational Risk vulnerability would involve a regulator who may consider adopting and maintaining a blacklist of Bitcoin addresses, whereby the bank taking a Bitcoin deposit would be obligated to search through the history of the deposited Bitcoins, and reject (and possibly report) the deposit. This may be created if the regulator has access to a registry linking Bitcoin addresses to real owners. There also exist tools to analyse the `taint', or the percentage of Bitcoins in a transaction that come from a known theft or scam [Moser, 2013]. Bitcoins involved in criminal transactions could also be `colored' [Rosenfeld, 2012], making it easy for a bank to identify them.

Such a requirement by the regulator may put a bank or financial institution, accepting deposits or transactions in Crypto currencies like bitcoin, who have inadequate controls, into a specific Operational Risk due to their failed action to check the address blacklist and therefore place them at risk of prosecution, e.g. for money laundering.

### 5.2.5   Handling of Sensitive Information

A fifth source of vulnerability that ought to be considered from an Operational Risk perspective for Crypto currencies like Bitcoin involves the handling of sensitive information. As noted previously, Bitcoin relies on public-key cryptography, an asymmetric key encryption scheme which is used for encrypting messages and verifying the originator of a message. Under this protocol, a user wishing to communicate, using a public-key cryptography scheme would have two keys: A public key that is available for everybody to access, as its name implies, and a private key that must be kept secret. Bitcoin uses the concept of digital signatures to ensure non-repudiation: that is, a third party can easily verify whether a particular signer has signed a message, using only information that is publicly available (the signer's private key).

This unique reliance on private keys, coupled with the irreversibility of payments in crypto currencies generates large operational risk potential losses specifically related to the following:

- Fraud and misappropriation of assets: If anyone gains access to a private key, he is able to create a transaction message and sign it as if they were the genuine owners, possibly transferring units of the currency to their own address.
- Loss due to processing errors: a loss process may arise for Bitcoin and other Virtual currencies relating to the fact that data entries for Bitcoin addresses may be inaccurately entered.
- Loss of electronic wallet due to the loss of technical support: in the case of Bitcoins, the most common storage account is known as an electronic `wallet', which stores the private/public key pairs for each of the user's Bitcoin addresses (each of which may hold a Bitcoin balance). The wallets may be stored on a user's computer or a mobile device, but may also be hosted online by a web service [Meiklejohn et. al., 2013].

If, in the process of controlling the wallets of its customers, the bank holding the Crypto currencies – or the client himself - misplaces, fails to secure or purposely tampered with such wallet files, access to the Bitcoins present in the affected accounts could be permanently lost. That is, while the Bitcoins themselves may continue to exist, and everyone in the blockchain could verify that they belong to certain accounts, without the private/public key pairs it would not be possible to access them, in order to use them in a transaction, for example. As an example, a Bitcoin service operator recently lost a significant amount of money due to hosting a wallet on non-persistent cloud storage, see details in [Barber, 2012].

Risk mitigation techniques for users are similar here to some of the classic methods of cyber security and data protection, including cold, offline, storage of digital wallets, storage on multiple devices which includes physical and digital, as well as solid encryption of private keys, strong passwords, and limited online transactions. Offline transaction signing and hardware wallets are a fast growing development for Crypto currencies for these reasons.

Moreover, the finite money supply decided for some currencies protocol like Bitcoins' make the loss irreversible as the money cannot be accessed, when losing the keys, nor regenerated, due to a limit in currency generation. Because of the finite money supply, the loss is permanent, and these Bitcoins are withdrawn from circulation permanently. A large event of this type or many events of this type due to negligence, electronic piracy or attack or internal fraud /damage to such clients e-wallets can cause a permanent problem for the Bitcoin money supply.

## 5.3 Exposures as Drivers of Operational Risk

Operational Risk exposures are intrinsic to the nature of any banking or financial business or product. Risk mitigation of exposures could obviously include the reduction of the assets at risk by diversification, when possible. Other solutions reside in impact reduction, contingency planning and business continuity management. In the following sections we highlight some Crypto currency specific risk exposures which may act as drivers for Operational Risk should banks or financial institutions choose to start accepting Crypto currency transactions.

### 5.3.1 Multiple Jurisdictions

In the context of tax non-compliance for banks accepting Crypto currencies, this can be a significant exposure since there is no easy way of figuring out the taxation rules regarding a currency that is universal and not associated with one country. In effect one would expect taxation to be relevant to the country in which the Crypto currency was stored and potentially receiving interest. However, due to the ease and difference in ability to move Crypto currencies such as Bitcoin around the world as a potential currency with less monitoring and regulation one would expect many issues related to tax avoidance to arise. If a bank or financial institution accepting transaction in Crypto currency was found to be complicit in this behaviour they would be exposed to large fines from regulators.

On its own this issue of registering clients details for a bank accepting deposits in Crypto currency accounts and e-wallets opens itself to a challenge as the taxation would need to be made on Crypto currency interest earned which could be either in the Crypto currency or in fiat currency, yet the taxation amount would need to be paid in fiat currency within a given jurisdiction. This makes it a particularly challenging task since several e-wallet providers and deposit taking institutions taking Crypto currencies are being proposed by multinational companies with potential operations in a range of jurisdictions. As such the accounts which are being discussed as international, will require to satisfy reporting requirements of particular account holders which will differ across different jurisdictions. This leaves the deposit taking intuition such as a bank taking Crypto currency exposed to Operational Risk losses in the form of regulatory fines if they fail to monitor and adhere to changes in each jurisdictions regulations.

### 5.3.2 Multiplicity of Micropayments

One of the benefits of Crypto currencies is the potential for consumers and businesses to make fractional payments for services that are repeated at very high volumes. The need for such small denominations is starting to become prevalent in some areas of financial transaction processing, see discussion in [Párhonyi et. al., 2005] and [Valdes-Benavides et. al.,2014]. Some special versions of currencies, like Crypto currencies have recently been developed by the Royal Canadian Mint, known as mintchips before they were abandoned, these e-money coins were created to allow fractional payments to be easily made in fractions of a cent. There is also the potential for Crypto currencies to be used in a similar manner. This is both an advantage for Crypto currencies over fiat currency with fixed smallest units eg. 1cent etc. and also a source of risk exposure that could act as a driver for Operational Risks. For instance, such mainstream use will require the ability to process massive volumes of many small transactions corresponding to fractions of cents, which may be undertaken by banking systems such as micropayment systems described in [Hernandez-Verme and Benevides]. Such micropayment systems have several aspects that differentiate them from standard electronic payment systems, primarily a major difference is that they typically process very large volumes of payments less than some small nominal amount, typically a fraction of a cent. This places a potential stress on processing systems to perform these tasks efficiently and accurately, and with such high volumes potentially being processed they could still produce substantial losses if not carefully considered from the perspective of risk controls and monitoring.

Therefore in banks accepting Crypto currencies with a view to such transaction processing there will be a very high risk that ultra-high frequency but very low consequence individual events can be perpetrated against a bank over extended periods of time before being detected, which on aggregate may create massive total losses, equivalent to a heavy tailed loss process with much lower frequency and significantly higher severity. These types of losses may be perpetrated by internal employees in the bank with sufficient knowledge of the micro-payment systems. These micro-payment systems due to their ultra-high frequency processing and minute payment amounts are susceptible to such attacks as already has been the case where massive losses were alleged to have been perpetrated against the Virtual currencies exchange in Tokyo, Mt.Gox, ultimately resulting in its closure in February 2014. The monitoring and reporting of these losses from an Operational Risk perspective will be a new challenge for both banks accepting such Crypto currency deposits and their regulators.

### 5.3.3 Hardware Reliance

The third Operation Risk exposure is related to specific hardware reliance that is required for processing and verification purposes in the Crypto currency networks. Whilst it is clearly the case that all modern banks and financial institutions require a large number of hardware components and an advanced IT infrastructure, the distinguishing feature of Crypto currency is that an entire network is required to remain in active, consistent and reliable operation. Whilst banks can control their IT infrastructure through resourcing adequately IT departments, this is not the case for the Crypto currency networks such as those associated to Bitcoin. The processing functions for blockchain verification and currency creation are external to any individual bank, making an external group of different geographically located entities critical to the reliable functioning of the Crypto currency network – including transaction processing and verification. This external hardware reliance creates a potentially massive Operational Risk exposure for any bank accepting processing of Crypto currencies and deposits.

There is also an additional hardware reliance related to storage of Crypto currency deposits. The vast majority of customer funds are held in cold storage, that is, offline, non-network connected storage, which minimizes the possibility that an attack will grant access to significant funds and thus make it significantly less susceptible to cyber-attacks and crime. For example, Coinbase keeps 98.8\% of funds in cold storage[19], the rest being in `hot wallets' connected to a computer networks or in transit. These wallets will be backed up and act as virtual bank vaults for Crypto currencies. If either of these cold or hot storage devices are destroyed or tampered with, there is a significant exposure to massive losses that the bank will obtain which is different to standard retail banking.

If a standard retail bank loses its database of client records, it would typically be backed up at least one alternative secure facility so that all accounts could be rapidly restored, and the same is possible for banks accepting Crypto currencies and customer wallet information. However, if one was able to tamper with a clients e-wallet and it is lost, corrupted or damaged, it is permanently lost and the bank would need to compensate the client for these losses. In addition, there is also the potential for massive losses from a hardware failure even if the Crypto currency e-wallets are intact, but instead the server holding the usernames and passwords for each client's e-wallet is lost. In this case, these would most likely be backed up and the losses would arise purely from a business disruption context as with standard banking practice.

### 5.3.4 Software Reliance

With regard to software there are clearly many risks that will each need to be monitored very carefully in a bank or financial institution accepting Crypto currencies. These include aspects of software rollout failures, protocol update disruptions etc. These are particularly relevant if the bank is taking deposits from a Crypto currency which has its protocol open to amendments or adjustments by different groups within the Crypto currency network, such as via the consensus voting or consensus mining frameworks discussed earlier. For instance it is possible that a bank will have to limit its operations due to changes in the Crypto currency specification: Bitcoin, and many other currencies, are consensus-based, and therefore a change in the protocol can be affected if the majority of the network agrees. Such changes may require updates in the clients' back-end operations and may cause disruption and ultimately will act as potential sources of Operational Risk exposures.

In addition, if the administrative component of the Crypto currency updates their software making old transaction protocols either void, open or susceptible to attack or significantly slower to process transactions, this would render significant business disruption losses. Losses of this type in banks holding Crypto currency have the potential to be substantial and should be monitored closely and understood by both regulators and banks considering accepting Crypto currency.

Software rollout issues also have the possibility of being more damaging when dealing with Crypto currency wallets, and if it is not carried out by a trusted party there is a possibility of incorporating malicious code in the release. Inadequate testing of production software may cause loss of access to wallet files upon the release of the software, and thus a permanent monetary loss due to transaction irreversibility. In addition, software architecture is a major consideration, as a system that is not scalable may fail from a sudden influx of customers, or a large increase in the number of transactions.

---

[19] http://antonopoulos.com/2014/02/25/coinbase-review/

## 6. Conclusion

From a regulatory perspective the European regulator recently published a position [EBA, 2014] discouraging financial institutions to accept and transact in Virtual currency before a regulatory framework is in place. We believe that the analysis performed in this paper can aid in understanding aspects important to be considered when developing this regulation and forming regulatory positions on banking sector interaction with Virtual and Crypto currency. The discussion developed in this paper should further inform these decisions form a risk management perspective, rather than the computer science and network design perspectives currently dominating the discussion on Virtual or Crypto currencies.

In this paper we develop a first attempt at understanding financial risk associated specifically to non-fiat Crypto currencies. In particular, we focus our analysis on Operational Risk components of such currencies under the Basel II/III banking regulation framework. In section 3 a novel development presenting a clear taxonomy of important risk management features of Virtual and Crypto currencies is developed. This facilitates a clearer understanding of attributes of these new forms of currency. We believe this can be of direct benefit to regulators, central banks and risk managers operating within financial institutions. We outline the distinction between Virtual (centralised) and Crypto (decentralised) currencies by developing a decomposition of features that distinguish Crypto currencies from other types of Virtual currencies. This is important for two reasons, the first reason is based on the fact that Crypto currencies are likely to be the most dominant form of Virtual currency in the real economy and therefore eventually in banking sectors as deposits, transactions and exchanges with fiat currency. Therefore, with a clear taxonomy of the wide variety of Virtual currency types, the second reason this is useful is that it allows one to focus effort on the most relevant sub-sets of such currencies and their features in order to develop an understanding of risk management vulnerabilities and exposures specifically related to the features of Crypto currency protocols and design.

In developing the taxonomy we focus on specific features of relevance to risk management analysis based on: issuance and generation of each of these forms of currency; monetary policy; administration of currency balances; storage and flows of crypto currency between individuals; exchange rates between Crypto and fiat currency; value generation; and currency area. From this taxonomy we develop a decomposition of Operational Risk types according to the key distinguishing features of Crypto currencies like Bitcoin.

Then in section 5 we develop a decomposition of Operational Risks for Crypto currencies first according to vulnerabilities specific to Crypto currencies which included: decentralised governance; peer-to-peer verifications of transactions; transaction irreversibility and handling of sensitive information. Then we develop a decomposition of Operational Risk exposures and risk drivers for such Crypto currencies including analysis of risk drivers such as multiple jurisdictions; multiplicity of micropayments; hardware and software reliance.

We showed from this analysis that anonymity, irreversibility of payments, private keys and decentralised network are amongst the features of Crypto currency that constitute vulnerabilities to a series of operational risks. Parallel to these vulnerabilities, software and hardware usage, and high volume of micropayments that characterise virtual and crypto currencies, leave them particularly exposed to some types of operational risk events.

Existing and advised ([Bitcoin.org](Bitcoin.org)) mitigating actions for users of Bitcoins relate to data protection and cyber security. They include cold, offline, storage of digital wallets, storage on multiple devices – physical and digital - , solid encryption of private keys and strong password, and limited online transactions. Offline transaction signing and hardware wallets are in a fast growing developments for these reasons.

On a more systemic aspect, the regulator may be concerned by the possible anonymity of Crypto currency operators, as the identifier of transactions is limited to public addresses and IP, without a clear possibility to know who operates from these addresses. Conversely, the peer-to-peer review systems requires a full traceability of operations, so that anyone can see the balance and the detail of every transactions operated by any address (testable on biteasy.com). This extensive transparency may not be convenient for every operator and is dramatically different from the principle of secrecy of the banking operations. A Bitcoin user that may not appreciate the full disclosure of his operation can install a transaction system that will generate a different address every time he executes a payment. This makes tracking more difficult for dishonest parties, but also for the regulator, facing there a potentially difficult challenge.

Recently, in the United States, [Brito and Dourado, 2014] praised the New York Department of Financial Services' (NYDFS) for outlining rules and regulations specific to virtual currencies (BitLicense proposal, 2014) but criticised some of the stipulations found in the document. The authors warn NY regulators not be more strict regarding the rules of Virtual currency transactions than for electronic payments, and not to treat young Virtual currency developers as established traditional financial institutions, at the risk of harming innovation and progress of the start-up like companies developing virtual currency systems. We endorse that view, provided that some methodological and technological securing mechanisms are appropriately developed. In fact, we intend that this paper further facilitates understanding of key aspects of vulnerabilities and risk drivers specific to such Crypto currencies being developed for real economies and entry into the banking sector en masse. We have focussed on what we believe to the be the most substantial risk class in terms of potential losses for financial institutions with inadequate risk modelling for such electronic currencies, namely Operational Risk.